| | |
|---|---|
| **Title** | **Fuel cell electrodes from organometallic platinum precursors: an easy atmospheric plasma approach** |
| **Authors** | Delphine Merche[1], Thierry Dufour[1], Joffrey Baneton[1], Giuseppe Caldarella[2], Vinciane Debaille[3], Nathalie Job[2], François Reniers[1] |
| **Affiliations** | [1] Service de Chimie Analytique et de chimie des Interfaces, Université Libre de Bruxelles, CP-255, 2 Bld du Triomphe, B-1050 Brussels, Belgium<br>[2] Department of Chemical Engineering – Nanomaterials, Catalysis, Electrochemistry, Université de Liège, Building B6a, Sart-Tilman, B-4000 Liège, Belgium<br>[3] Laboratoire G-Time, Université Libre de Bruxelles, CP 160/02, 50 Av F.D. Roosevelt, B-1050 Brussels, Belgium |
| **Ref.** | Plasma processes and polymers, 2016, Vol. 13, Issue 1, 91-104 |
| **DOI** | http://dx.doi.org/10.1002/ppap.201500157 |
| **Abstract** | An organometallic powder (platinum (II) acetylacetonate) is decomposed in the post-discharge of an atmospheric RF plasma torch to deposit Pt nanoparticles on carbon black supports. The resulting nanohybrid materials are characterized by FEG-SEM and XPS techniques to highlight their high content in Pt, their oxidation degree, and the dispersion of the Pt nanoparticles on the substrate. ICP-MS and electrochemical characterizations in a single fuel cell (cyclic voltammetry, dynamic polarization curves) are also performed on electrodes realized by treating the powder mixture overlaid on gas diffusion layers. The comparison of the catalytic activity and the Pt loading with commercially available electrodes shows the great potential of this simple innovative, fast, and robust deposition method. |

# 1. Introduction

The deposition of noble metal nanoparticles on carbon supports (including carbon nanotubes, graphite, carbon black, activated carbon, …) has been the subject of a great interest over the last few years, due to potential applications in sensing[1–3] and actuation,[4] but also in catalysis with the emergence of components such as proton exchange membrane fuel cells (PEMFC).[5,6] In all cases, carbon matrices have encountered a great success as they provide both a mechanical support and a continuous electronic medium. The introduction of platinum catalysts supported on carbon black is known to decrease the platinum loading to less than $0.1 mg_{Pt}.cm^{-2}$, while preserving the fuel cell performances.[7–10]

Several methods for the synthesis of nanoparticles on carbon supports were reported in the literature: wet impregnation,[11] chemical[12] or electrochemical[13] precipitation, colloidal techniques,[14] etc. Although these techniques proved to be more or less effective in the synthesis of nanoparticles on carbon supports, they all present consistent drawbacks associated with the complexity of the experimental protocols.

Alternatively, more environmental friendly techniques have arisen to produce metal nanoparticles in the gas phase by (low pressure) chemical vapor deposition (LP) CVD,[15,16] flame pyrolysis,[17] laser ablation,[18,19] or (low pressure) physical vapor deposition (LP) PVD, e.g., magnetron sputtering.[20,21] Magnetron sputtering allows the synthesis of catalytic layers of low platinum loading and presenting high power densities, thanks to a catalyst deposition occurring mainly on the support (sub)-surface. The platinum profile distribution closely depends on the process variables. Moreover, the use of various targets (including alloys) enables the elaboration of multiple compounds (co-deposition Pt/C, Pt/Ru, alternate deposition C/Pt, …) to realize catalytic layers on Nafion by plasma with an appropriate distribution of the catalyst. However, the major drawback of this technique is the mandatory of operating under ultra-high vacuum. Nanoparticles can also be synthesized from vapors of organometallic compounds based on a plasma scheme by metal-organic chemical vapor deposition (MOCVD). In this process, the precursors are often dissociated in the plasma volume to nucleate particles in the gas phase. In this case, the process requires the sublimation of the metallic powder and an extremely short residence time (1ms) of the vapor in the atmospheric microdischarge for a very high plasma power.[22,23]

The anchoring of metallic nanoparticles on a support requires the presence of anchoring sites. Many studies concerning the plasma activation of carbon supports were conducted to improve the interfacial interactions between the support and the deposited metal nanoparticles.[24,25] The activation, allowing the anchoring of smaller clusters more homogeneously distributed and enhancing the stability of the deposited compound is often performed with oxygen or argon plasma.[24] It is usually based on the grafting of oxidized species but also on the formation of structural surface defects (achieved by ion or electron bombardment). The density of the chemical and surface defects plays an important role on the dispersion and the size of the clusters. This activation is usually followed by a step of metallic nanoparticle deposition. For example, Bittencourt et al.[26] evidenced the formation of C—O—Pt bonds by realizing an oxygen plasma treatment followed by a ultra-high vacuum platinum evaporation.

Previous articles from our group[27–30] reported the deposition of metal nanoparticles (Rh, Au, Pt, . . .) on carbon substrates (carbon nanotubes, HOPG, glassy carbon) by spraying a colloidal solution in the flowing post-discharge of a radio-frequency (RF) atmospheric plasma torch. The use of an atmospheric plasma torch enables to obtain the surface activation and the surface decoration in ‘‘one single step,’’ contrarily to most of the techniques dedicated to the deposition of metallic nanoparticles on carbon supports. Although the nebulization of a platinum colloidal solution within the post-discharge allowed a homogeneous dispersion of the anchored nanoparticles but also a size similar to that of the nanoparticles in the colloidal solution, the prior synthesis of the appropriate Pt colloid solution is a time-consuming and delicate process based on organic reactions. Preparing these colloids ‘‘on site’’ is therefore another complicated step, and purchasing such a solution leads also to a significant financial investment (compared to the real amount of Pt present in the solution) not suitable for applications requiring the use of large quantities of nanoparticles. For this reason, we propose to substitute the colloidal solution by an organometallic source. In the scope of this article, Pt nanoparticles from platinum (II) acetylacetonate [$Pt(acac)_2$] were synthesized and deposited on activated carbon black (CB) powder by means of an RF atmospheric plasma torch supplied with argon in an open air environment.

In addition to a fundamental study of the effect of several synthesis variables on the platinum content and on the platinum oxidation states for the powder mixture pressed on a copper tape, electrodes for PEMFC applications were also realized by treating the powder mixture overlaid on gas diffusion layers (GDL). The electrodes were tested in a mono-fuel cell and compared with commercial electrodes. This technique is very simple, fast, and robust. It also presents substantial economic advantages, such as the non-requirement of a vacuum system and the possibility to easily implement this process in a continuous production line. Moreover, our process operates at atmospheric pressure, at temperatures close to ambient temperature, and avoids sublimation of the precursor or its nebulization into the (post)-discharge.

The influence of the treatment time, the gap (distance between the sample and the torch), the ambient atmosphere, and the temperature on the mixtures properties were studied by XPS regarding the surface composition and the oxidation state of the platinum nanoparticles. The metallic nanoparticles were observed by a FEG-SEM to highlight their repartition on the surface. Electrochemical characterizations (voltammetry and dynamical polarization curves) of selected GDL electrodes decorated by Pt nanoparticles synthesized using this original approach were realized to determine the electrochemical active surface and the catalytic activity of the platinum particles as a function of their amount in the samples (determined by ICP-MS).

# 2. Experimental Section

## 2.1. Plasma Torch

The post-discharge was generated by an RF plasma torch (Atomflo-250DfromSurfx Technologies) using argon as the vector gas.[31] The plasma torch, illustrated in Figure 1, consisted of two closely spaced circular aluminum electrodes perforated to flush the argon gas at a flow rate of 30 L.min$^{-1}$. For these experiments, the plasma torch could be used in two configurations: (i) the ‘‘open configuration,’’ where the powder was pressed on a copper tape or overlaid on the GDL and stuck on a glass slide (Figure 1a); and (ii) the ‘‘confined configuration,’’ where the influence of the ambient air was prevented by pressing the powder on a copper tape fixed at the bottom of a crucible placed against the plasma torch (Figure 1b).

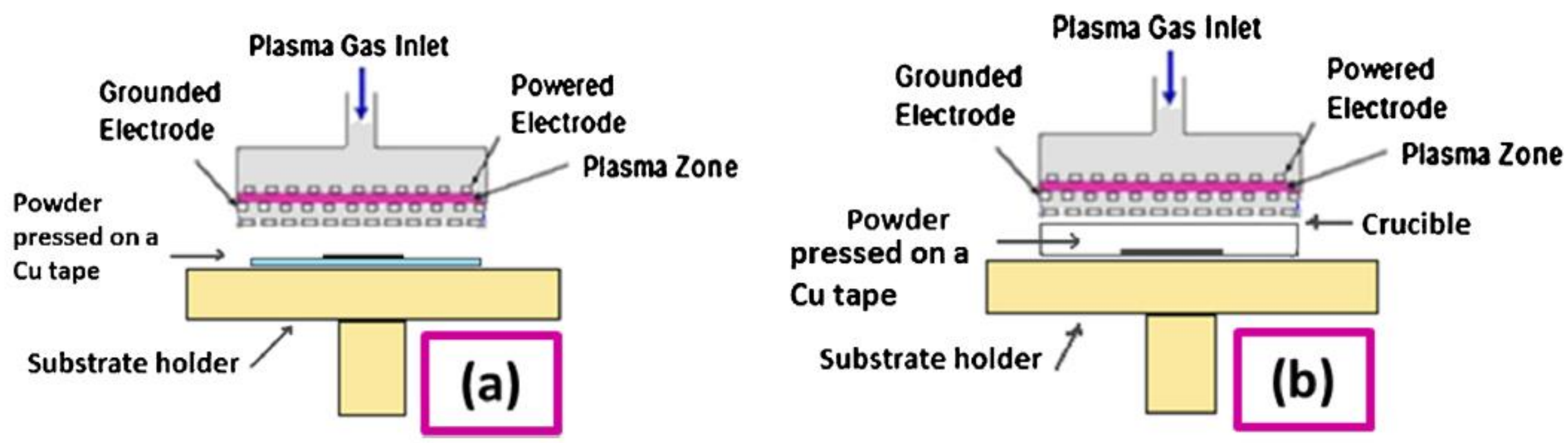

***Figure 1. Scheme of the RF plasma torch (Atomflo-250D SurfX) for two configurations:***
***(a) the open air configuration where the substrate is fixed on a glass slide, (b) the confined configuration.***

## 2.2. Materials

The carbon black ENSACO 290 support was provided by TIMCAL Graphite and Carbon. Its surface area was ranging from 226 to 236m$^2$.g$^{-1}$ and the micropore volume was comprised between 0.055 and 0.065 cm$^3$.g$^{-1}$. Platinum acetylacetonate [$(C_5H_7O_2)_2^-$ $Pt^{2+}$], named ‘‘[Pt(acac)$_2$]’’ in this paper, was purchased from Sigma–Aldrich with a purity of 99.99%. The gas diffusion layers (GDLs) were purchased from Paxitech. These GDLs consist of a support based on a carbon felt treated with a fluoropolymer (PTFE, 10% in mass) for the water management (water removal at the cathode and water retaining at the anode to prevent the membrane from drying-out) and contain a carbon microporous layer made of carbon black (CB).[32] The membrane-electrode assemblies realized for the single-fuel cells tests are composed of (i) a Nafion membrane NR-212 (50µm thick); (ii) an anode consisting of catalyst on a carbon microporous layer (Pt/C of 70wt% (HSC)) supported on a Freudenberg GDL and characterized by a Pt loading of 0.5 mg$_{Pt}$.cm$^{-2}$ (H2315 I2 C6); and (iii) a cathode. The cathode is either the plasma electrode or a commercial electrode named in this study ‘‘reference electrode,’’ presenting the same loading and mass ratio as for the anodic size. It was provided by Paxitech (GDE-225-70-0.5 metal/air, teflon: 10 wt%).

## 2.3. Preparation of the Powder Mixture

The mixture of organometallic precursor with the carbon powder support was achieved as follows for a preparation of 75 mg:

(i) 25mg of [Pt(acac)$_2$] were dissolved in acetone (5 mL) during 15 min in a ultrasonic bath. Then, 50mg of carbon black powder were added under ultra-sonication during 5 min.

(ii) The solvent was evaporated from the mixture in a stove at 353 K.

(iii) The resulting dried mixture, with a platinum weight ratio ($W_{Pt}$) of 16 wt%, was manually crushed and a thin coating of the powder (0.8 mg) was pressed onto a copper tape (1cm$^2$).

(iv) The tape with the thin coating was (a) either fixed on a glass slide placed at a gap (i.e., the distance separating the plasma torch from the slide) of 3 or 10mm or (b) placed in a crucible 10mm in height (of the same diameter than the diameter of the active area of the torch, i.e., 2.5 cm) positioned against the plasma torch. The glass slide or the glass crucible lied on a stainless steel substrate holder.

## 2.4. Interest to Press the Mixture Powder

Pressing the powder on a metallic tape presented three main advantages: (i) the powder was not blown away by the argon flow rate; (ii) the thickness of the pressed powder mixture layer was thicker than the part of this layer which was catalyzed, thanks to the plasma treatment. The latter was itself larger than the XPS probing depth (10 nm). Therefore, the XPS measurements were carried out on the powder treated by the post-discharge only; (iii) stirring of the powder was avoided.

As the plasma torch can treat the surface of the powder exposed to the post-discharge only, the bulk of the layer remains mostly unaltered. However, according to the work of Antoine et al.[33] dealing with the influence of platinum localization in the active layer of a porous electrode, the performances of electrodes could be optimized provided that the Pt nanoparticles were mostly present close to the protonic membrane. Indeed, the platinum nanoparticles can be active only if they are in contact at the same time with (i) the combustible/fuel, (ii) the carbon black powder, and (iii) the electrolyte solution or membrane (triple contact point).

## 2.5. Powder Mixture Overlaid on Gas Diffusion Layers

Electrodes were also achieved by spreading an ink, constituted of 4mg of the powder mixture ($W_{Pt}$=16 wt%) and a few droplets of methanol on GDLs of 2 cm$^2$. The resulting electrodes were analyzed by XPS, SEM, and ICP-MS. For electrochemical characterizations (tests in a single-cell), GDLs of 25cm$^2$ were covered by 50mg of powder. The plasma (''open configuration,'' gap¼3mm) was operated on a 5cm$^2$ area and was repeated five times to cover the 25cm$^2$ of the GDL. The results were compared with those of a single-fuel cell containing a commercial electrode (''reference electrode'').

## 2.6. Characterizations

XPS analysis of the plasma-treated samples were carried out with a Physical Electronics PHI-5600 system equipped with a concentric spherical analyzer and a 16 channeltron plate. Spectra were acquired using the Mg anode (1253.6 eV) operating at 150 W. Survey scans were acquired on the 0–1000 eV range at a 187.85 eV pass-energy with a five-scan accumulation (time/step: 50ms, eV/step: 0.8). The XPS spectra for elements of interest were acquired at a 23.5 eV pass energy with an accumulation of ten scans (time/step: 50ms, eV/step: 0.05). The CASA XPS software was used to calculate the relative elemental composition. It was calculated on the survey scan by measuring the peaks areas after subtraction of a Shirley background. The following sensitivity coefficients were used: C=0.205, N=0.38, O=0.63, and Pt=3.06 (instead of 1.75 corresponding to the singlet $Pt4f_{7/2}$) as referred in the literature.[34] Indeed, in order to take into account all the platinum components and avoid neglecting the Pt (+4) and Pt (+2) components, the whole spectral envelope of Pt on the survey scan, corresponding to the spin-orbit doublet ($Pt4f_{7/2}$ and $Pt4f_{5/2}$), was considered. The sensitivity coefficient was calculated by considering the ratio of these two peak singlets areas (ratio 4f7/2 on $4f_{5/2}$=4/3 from the literature).[34,35] The XPS compositions were only used to determine the relative effect of the plasma variables on the amount of Pt deposited on the surface and not for absolute quantification. The C 1s and Pt 4f XPS peaks were decomposed into several components of same FWHM, through convoluting Gaussian and Lorentzian profiles with respect to a 30:70 ratio. The component assignments and binding energies were based on the literature.[35,36] The binding energy calibration of the Pt 4f XPS peak was corrected with respect to the C1s spectra (C—C component of the carbon black at 284.6 eV).[37]

Field electron gun-scanning electron microscopy (FEG-SEM) was used to observe the morphology of the catalyst powder and the size of the nanoparticles deposited on the carbon support. The FEG-SEM images were obtained using a JEOL JSM-7000F instrument equipped with an energy dispersion X-ray spectrometer (EDX, JED-2300F) which could be performed on a submicron analysis volume. The equipment was operated at 10 kV accelerating voltage. The images obtained by the secondary electron emission showed the morphology and the roughness while those obtained by the backscattered electrons (COMPO detector) highlighted the presence of metallic particles (effect of the atomic number).

An agilent 7700 inductively coupled plasma mass spectrometer (ICP-MS) equipped with a He collision cell was used to determine the absolute quantity of metallic platinum per area unit of GDL. The powder was collected by totally scraping the plasma-electrode. Then it was dissolved in a 3:1 mixture of concentrated sub boiled nitric (14N) and hydrochloric (6N) acids for 48 h on a hot plate at 393 K. After evaporation, samples were redissolved in 14N $HNO_3$, again for 48 h at 393 K. After evaporation, they were centrifuged in 0.05N $HNO_3$ and diluted for analysis. The concentration of Pt was calculated in function of a calibration curve (based on 10, 50, 100, 500, and 1000 ppb standards in the presence of indium as internal standard). The total reproducibility of Pt analysis based on duplicate measurements performed on distinct-diluted solutions of three samples is better than 2.5% (2 RSD).

Electrochemical characterizations were performed on the electrodes synthesized on GDL. The assembly of the fuel cell (single cell test fixture of 25$cm^2$ with graphite bipolar plates) containing the plasma electrode consists in:

(i) Spraying an ionomer (Nafion) solution onto the electrodes to obtain a concentration around 0.6–1$mg.cm^{-2}$. Then, the system was dried in an oven at 333 K.

(ii) Combining all the independent layers provided by Paxitech to form a complete fuel cell membrane-electrodes assembly (Nafion 212, commercial anode from Freudenberg (of 0.5 $mg_{Pt}.cm^{-2}$), plasma-treated Pt on GDL as cathode and sealing gaskets), and hot-pressing the complete assembly at 403K during 210 s (total force of 30 kN).

(iii) Activating the device supplied by gas channels ($H_2$ flow rate = 80 $mL.min^{-1}$ and $O_2$ flow rate = 350 $mL.min^{-1}$) under thermostatic control (343 K) until electrical stabilization (measured by impedance spectroscopy).

The electrochemical characterization of the single fuel cell containing the plasma-electrode consisted in determining (i) the electrochemical surface area (ECSA) and (ii) the performance of the fuel cell in terms of polarization curves. The potentiostat is a Biologic Model SP-150 combined with a booster 80A. The electrochemical surface area (ECSA) was determined by cyclic voltammetric analysis (CV). The voltage was scanned between 0.04 and 1.2V with a scan rate of 20mV.$s^{-1}$ (hydrogen flow rate: 80 mL.$min^{-1}$, 100% humid air: 350 mL.$min^{-1}$ ). The ECSA ($m^2.g_{Pt}^{-1}$) is given by Equation (1): ECSA = Q./(L.$\Gamma$) where Q (C.$m^{-2}$) is the charge density corresponding to hydrogen desorption from platinum obtained from the CV (voltage range around 0.1 and 0.3 V), L ($g_{Pt}.m^{-2}$) is the platinum loading of the electrode obtained from ICP-MS and $\Gamma$ (C.$m^{-2}$) is the charge required to reduce a monolayer of protons on polycrystalline platinum.[38,39] It is considered equal to 210.$10^{-2}$ C.$m^{-2}$.

Performances of the fuel cells were estimated using dynamic polarization curves at 70°C. The voltages range is comprised between OCP and 0.2V while the scan rate is 20 mV.$s^{-1}$ ($H_2$ flow rate: 250 mL.$min^{-1}$, 100% humid air: 450, 750, and 1000 mL.$min^{-1}$). Voltage losses due to redox, ohmic resistance and transport phenomena directly influence the shape of the curves, allowing to estimate the importance and limitations of gas diffusion, ohmic resistance (and so contacts between the different layers of the fuel cell) and the catalytic activity of platinum particles.[40]

# 3. Results and Discussion

## 3.1. Influence of the Plasma Treatment Time

The optimal treatment time of the powder mixture to produce a significant amount of Pt nanoparticles on the substrate while minimizing the quantity of plasmagen gas (necessary for potential industrial applications) was evaluated. This study was also performed to investigate whether the exposure time could influence the oxidation state of the nanoparticles. Each sample (pressed powder mixture) was exposed to an argon post-discharge with a flow rate of 30 L.$min^{-1}$ delivered by the RF plasma torch powered at 60 W. The substrates were placed 3mm downstream from the plasma torch (''open configuration'' in Figure 1a) for various exposure times: 0, 30, 60, 150, 300, and 600 s.

A study of the elemental surface composition was performed by XPS, as illustrated in Figure 2 in which Pt 4f, C 1s, N 1s, and O 1s peaks are evidenced at 71, 285, 400, and 533 eV, respectively. The intensity (counts) of the Pt 4f peak clearly increases with the exposure time. Table 1 reports the relative surface composition of each exposed sample. The composition of the native mixture is close to that of the native carbon black (93.4$\pm$1.8% C and 6.6$\pm$1.8% O). The native powder mixture presents an amount of Pt as low as 0.3% while it increases to 6.1% after a plasma treatment of 30 s and up to 13.3% after 600 s of plasma exposure. A small amount of nitrogen is detected on the sample surface. The O1s peak detected on the untreated powder mixture (6.4%) may reflect a slight and native oxidation of raw carbon black powder but also (to a lower extent) the presence of the four O atoms in the formula of the [$Pt(acac)_2$]. After a plasma exposure of 600 s, the oxygen content increases to 17.1%. The observed surface oxidation could result from the sample/post-discharge interaction and/or with the ambient air after plasma treatment.

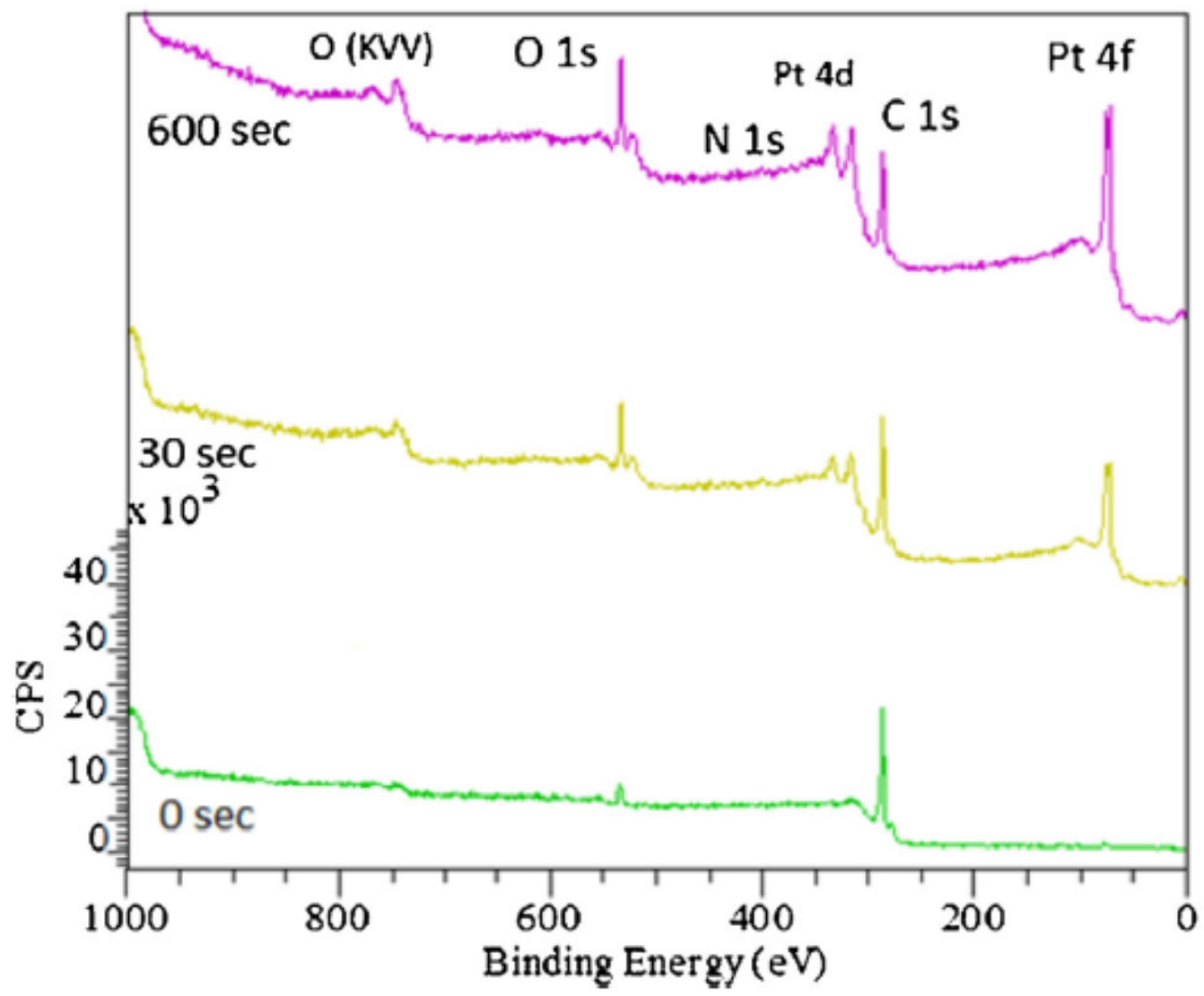

| Time [s] | %Pt | %C | %N | %O |
|---|---|---|---|---|
| 0 | 0.3 ± 0.1 | 93.4 ± 1.2 | 0 | 6.3 ± 1.3 |
| 30 | 6.1 ± 0.8 | 77.4 ± 1.1 | 2.3 ± 1.5 | 14.2 ± 1.0 |
| 60 | 6.9 ± 0.7 | 75.4 ± 1.2 | 2.7 ± 0.6 | 15.0 ± 1.3 |
| 150 | 8.3 ± 0.7 | 74.2 ± 1.5 | 2.8 ± 0.3 | 14.7 ± 0.9 |
| 300 | 9.9 ± 2.3 | 71.5 ± 3.1 | 0.9 ± 0.9 | 17.7 ± 2.4 |
| 600 | 13.3 ± 3.0 | 68.1 ± 1.4 | 1.5 ± 1.5 | 17.1 ± 1.8 |

***Table 1. Atomic surface composition of the CB/[Pt(acac)$_2$] mixture treated in the post-discharge of the RF plasma torch supplied in Ar, as a function of the exposure time (determined by XPS).***

***Figure 2. Survey scan spectra of the CB/[Pt(acac)$_2$] untreated mixture, the mixture treated during 30 and 600 s (P=60 W, Ar flow rate=30 L.min$^{-1}$, gap=3 mm).***

Although the mixture was characterized by a $W_{Pt}$ of 16%, the XPS scarcely detected the presence of Pt on the untreated samples. This could be due to the small depth analysis of the XPS technique (10nm) in comparison to the grain size of the carbon black (50nm of diameter and forming micro-agglomerates). Moreover, most of the organometallic powder could be dispersed within the interstices of the porous carbon black powder before any plasma treatment. Several assumptions could explain the increase in the Pt content induced by the plasma treatment: (i) the functionalization and the formation of surface defects on the carbon grains to generate anchoring sites for the platinum nucleation; the treatment of the native carbon black powder lead to 17.8±1.1% of oxygen after 300 s of plasma treatment (increase of the C=O and COOR components as in agreement with the literature);[37] (ii) the release in the gas phase of carbon-based oxidized fragments issued from the degradation of the organometallic powder by the energetic species of the flowing post-discharge and from thermal effects; (iii) a massive plasma-induced oxidative etching of the carbon black support particles; (iv) a thin topmost layer of non-catalyzed carbon grains that could be blown away by the argon flow.

The oxidation state of Pt was also investigated by deconvoluting the high-resolution Pt 4f peak, based on a method proposed by Croy.[35] The deconvoluted spectra are represented in Figure 3. The two spin-orbits singlets Pt $4f_{7/2}$ and Pt $4f_{5/2}$ are separated by an energetic gap of 3.3 eV for an area ratio Pt $4f_{7/2}$/Pt $4f_{5/2}$ of 4/3. According to the fitting of the Pt 4f XPS peaks and as reported in Table 2, no Pt under metallic form was detected for the untreated mixture of CB/[(Pt-(acac)$_2$] while the most intense Pt component of the plasma-treated sample was Pt (0). 70% of platinum was found to be under its metallic form after 30 s of plasma treatment and 78% after 600 s. The Pt (0) component increases with the treatment time while the Pt (+2) component decreases. This latter component might be attributed to the undecomposed [Pt(acac)$_2$], to the surface oxidation of the platinum nanoparticles and/or to the nanoparticle-support interactions (Pt—O—C bond). (The small differences in relative compositions between the analysis of the Pt $4f_{7/2}$ and the Pt $4f_{5/2}$ signals are mostly within the error margins of the instrument and could be due to the fitting parameters as well as to the background subtraction).

| Transition | Position [eV] | Assignment | % at 0 s | % at 30 s | % at 600 s |
|---|---|---|---|---|---|
| Pt $4f_{7/2}$ | 71.0 | Pt (0) | 0 | 70.0 | 78.1 |
| | 72.4 | Pt (+2) | 58.6 | 22.5 | 15.5 |
| | 74.6 | Pt (+4) | 41.4 | 7.5 | 6.4 |
| Pt $4f_{5/2}$ | 74.3 | Pt (0) | 0 | 70.7 | 76.8 |
| | 75.7 | Pt (+2) | 61.1 | 22.7 | 16.8 |
| | 77.9 | Pt (+4) | 38.9 | 6.6 | 6.4 |
| C 1s | 284.6 | C—C | 83.6 | 73.4 | 70.3 |
| | 286.5 | C—O—R | 12.9 | 16.9 | 17.9 |
| | 288.0 | RR′C=O | 1.7 | 5.3 | 7.2 |
| | 289.2 | COOH, COOR | 1.8 | 4.4 | 4.6 |

***Table 2. Chemical components of the Pt4f and C1s XPS spectra, for untreated samples, and samples treated during 30 and 600 s.***

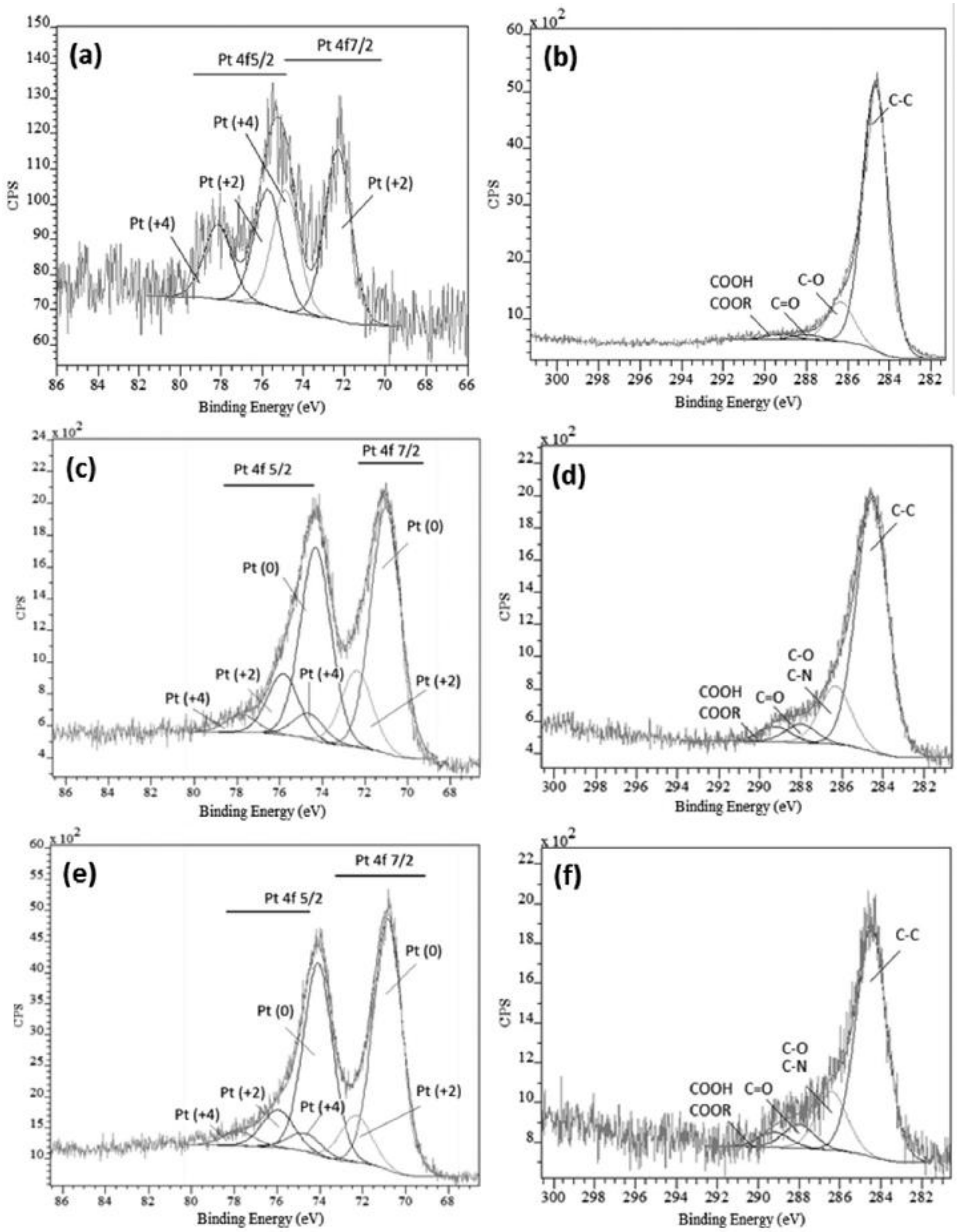

***Figure 3. Pt 4f and C 1s XPS peaks for (a, b) CB/[Pt(acac)$_2$] untreated mixture (0 s), (c, d) mixture treated for 30 s, and (e, f) mixture treated for 600 s.***

The peak at 284.6 eV is attributed to the C—C component (C $sp^2$ and C $sp^3$ orbital hybridizations). The peak at 286.5 eV is mainly assigned to ether functionalities. The 288.0 eV component, related to the ketones, increased with the treatment time. The 289.2 eV peak corresponds to esters and carboxylic acids functionalities. The increase of the O concentration in XPS is consistent with the increase of the C—O components in the C 1s peak. This partial oxidation of carbon is thought to favor the fixation of Pt particles on carbon substrates.[25,41]

## 3.2. Influence of the Gap and the Environment

The gap is a variable of great interest for any plasma treatment achieved with a flowing post-discharge: too small, it provokes undesired temperature effects that can damage the substrate while, too large, it induces a decrease of the treatment efficiency.[42] For instance, when the gap increases from 3to10mmina pure Ar post-discharge, the Pt content decreases from 9.9 to 3.5%. Moreover at this distance, no platinum was present under metallic form as shown in Figure 4.

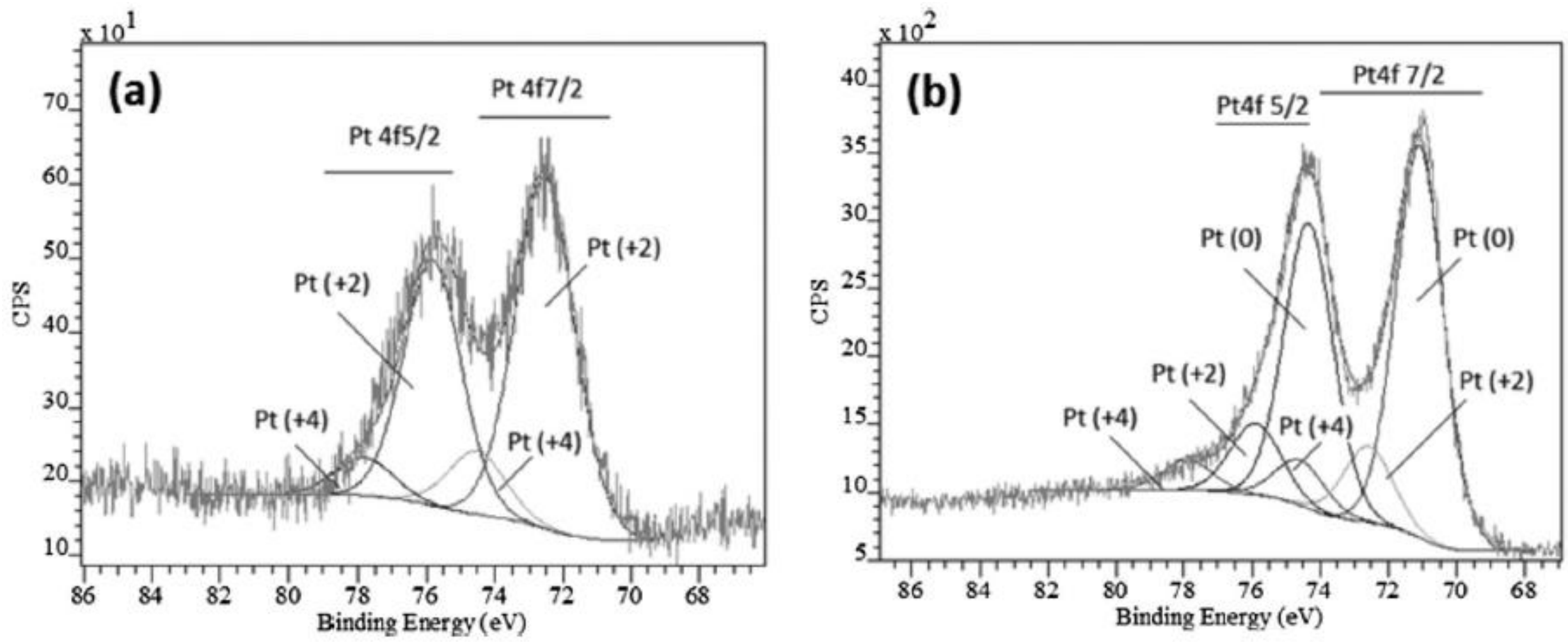

***Figure 4.** Comparison between Pt 4f spectra for a plasma treatment of the CB/[Pt(acac)$_2$] mixture during 300 s (60 W, 30 L.min$^{-1}$ Ar, gap=10mm) (a) in the ''open configuration'' and (b) in the ''confined configuration.''*

According to optical emission spectroscopy (OES) measurements performed for a gap of 3mm, the species from the post-discharge are mostly long-lived species, namely molecules, radicals, and excited atomic species.[43] At 10mm, the overall emission of the excited Ar species also drops considerably due to collisions with other gaseous molecules, thus explaining a less significant degradation of the [Pt(acac)$_2$] organic component and an enhance of the platinum oxidation. Such a configuration with a gap of 10 mm is therefore not interesting. Nevertheless, in the case of the confined configuration (mixture on a tape stuck on a crucible placed against the torch), the decrease in the Pt content was not observed. Indeed, the Pt content reached 11.4% for a gap of 10mm. Moreover, the confined configuration minimizes the ambient pollution, preventing oxygen species of the environment from reacting with the excited species (13.1% O compared to 18.2% in the open configuration for the same gap). Therefore, the Pt is scarcely oxidized (75% of Pt is under metallic form, cf. Figure 4 and Table 3) and even less than for a gap of 3mm with the open configuration (71.4% Pt (0)).

| Gases | Gap [mm] | Configuration | %Pt | %C | %N | %O | Pt (0) |
|---|---|---|---|---|---|---|---|
| / | / | / | $0.3 \pm 0.1$ | $93.4 \pm 1.2$ | 0 | $6.3 \pm 1.3$ | – |
| Ar | 3 | Open | $9.9 \pm 2.3$ | $71.5 \pm 3.1$ | $0.9 \pm 0.9$ | $17.7 \pm 2.4$ | 71.4 |
| Ar | 10 | Open | $3.5 \pm 1.0$ | $77.8 \pm 1.7$ | $0.5 \pm 0.5$ | $18.2 \pm 1.7$ | 0 |
| Ar | 10 | Confined | $11.4 \pm 1.4$ | $75.0 \pm 2.0$ | $0.4 \pm 0.4$ | $13.1 \pm 1.3$ | 74.7 |

***Table 3. XPS relative composition and metallic platinum content (Pt (0)) of the CB/[Pt(acac)$_2$] mixture treated by the Ar post-discharge for various gaps and configurations.***

The ‘‘confined configuration’’ (as far as 10 mm), and the ‘‘open configuration’’ for a gap of 3mm present the best results. The advantage of the latter is the ability to treat larger carbon substrate areas. Therefore, this configuration was adopted to synthesize plasma-electrodes for electrochemical characterizations in a single fuel cell.

## 3.3. Influence of the Temperature

In order to favor the degradation of the [Pt(acac)$_2$] organic component by a synergetic effect of the temperature during the plasma process, the influence of the temperature on the elemental composition was investigated. The samples (mixture CB/[Pt(acac)$_2$]) pressed on the copper tape were stuck on a glass slide placed on a hot plate. They were heated 2 min before the plasma treatment and then exposed to the argon post-discharge for 300 s (RF power of 60W and gap of 3 mm). The elemental surface composition of the samples is presented in Table 4 for various temperatures. The temperature of 333 K corresponds to a surface heating induced by the post-discharge only (hot plate off) while the other temperatures resulted from the heating of both the post-discharge and the hot plate.

| Surface temperature | Plasma torch | %Pt | %C | %N | %O | %Cu |
|---|---|---|---|---|---|---|
| Native | OFF | $0.3 \pm 0.1$ | $93.4 \pm 1.2$ | 0 | $6.3 \pm 1.3$ | 0 |
| 333 K | ON | $9.9 \pm 2.3$ | $71.5 \pm 3.1$ | $0.9 \pm 0.9$ | $17.7 \pm 2.4$ | 0 |
| 353 K | ON | $13.0 \pm 2.3$ | $65.9 \pm 4.1$ | $1.9 \pm 1.4$ | $18.9 \pm 2.0$ | $0.3 \pm 0.3$ |
| 383 K | ON | $18.0 \pm 2.4$ | $64.4 \pm 4.8$ | $0.6 \pm 0.6$ | $16.7 \pm 3.0$ | $0.3 \pm 0.3$ |
| | OFF | $0.2 \pm 0.0$ | $90.2 \pm 0.4$ | 0 | $9.6 \pm 0.5$ | 0 |
| 423 K | ON | $15.3 \pm 1.4$ | $65.3 \pm 2.0$ | 0 | $18.5 \pm 1.4$ | $0.9 \pm 0.5$ |
| 473 K | ON | $6.5 \pm 2.6$ | $72.0 \pm 1.9$ | 0 | $19.5 \pm 0.8$ | $2.0 \pm 0.5$ |
| 543 K | ON | $<0.7 \pm 0.1$ | $79.7 \pm 1.9$ | 0 | $18.5 \pm 1.7$ | $1.1 \pm 0.4$ |
| | OFF | 0 | $85.7 \pm 1.7$ | 0 | $14.3 \pm 1.7$ | 0 |

***Table 4. Elemental surface composition evidenced by XPS for several surface temperatures (open configuration, 60 W, 300 s of plasma treatment, 30 L.min$^{-1}$ Ar).***

For a surface temperature ranging from 333 to 383 K, the Pt content increased from 9.9 to 18.0%, while it decreased to less than 0.7% for a temperature as high as 543 K. The Pt content for this latter temperature could even be overestimated since the Cu 3p and Pt 4f peaks appear at very close binding energies. Such a low content probably results from the sublimation of Pt(acac)$_2$, which ranges, according to the literature, between 513 and 533 K.[44]

According to the elemental composition of the mixture heated at 383 and 543K without plasma treatment, Pt remains almost undetected by XPS (less than 0.2%). Moreover, the oxygen content is lower in the absence of treatment. Thus, one can conclude that the plasma favors the decomposition of the organometallic and the generation of nucleation sites (oxygenated functionalities or defects) for depositing metallic nanoparticles on the carbon powder. The decomposition of the organometallic could be assumed to occur below 533 K in the post-discharge due to the interaction between its energetic species ($Ar^+$, Ar*, electrons, etc.) and the [$Pt(acac)_2$] powder. Therefore, organic ligands could be degraded and metallic nanoparticles synthesized below the decomposition temperature (533 K).

The platinum nanoparticles synthesized at room temperature on carbon black powder for a gap of 3mm (300 s) by the RF plasma torch were observed by SEM, as shown in Figure 5: pictures a, b, and c were obtained from backscattered electrons and pictured was recorded with secondary electrons in order to observe the topography, the morphology, and the size of the carbon nanoparticles.

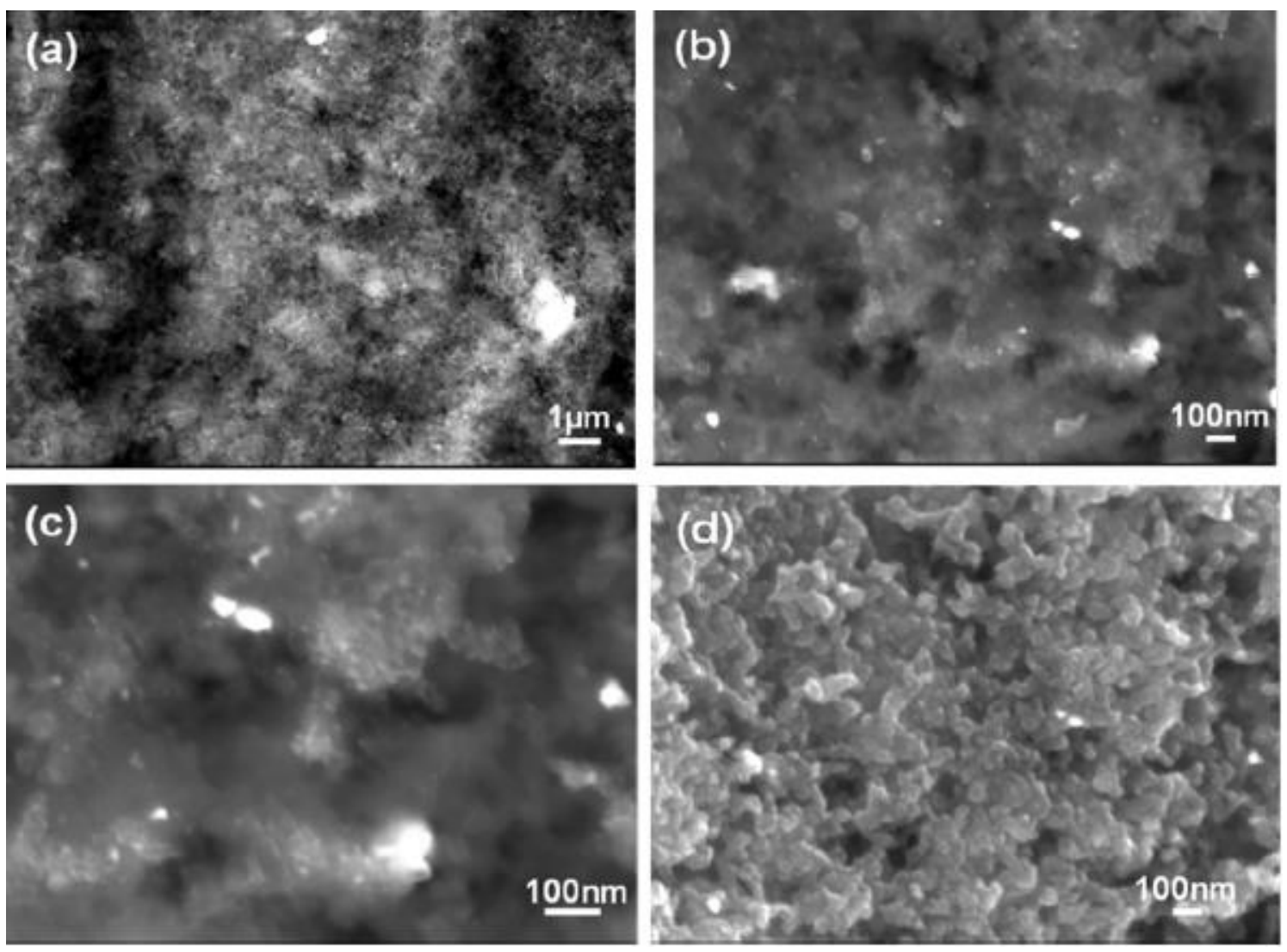

***Figure 5. SEM pictures of the CB/[Pt(acac)$_2$] mixture exposed during 300 s to the flowing post-discharge (60 W, gap = 3 mm, Ar flow rate = 30 L/min): (a–c) backscattered electrons, (d) secondary electrons.***

The carbon nanoparticles form micrometric aggregates presenting a bean shape with a typical diameter of 50 nm. Due to the effect of the atomic number, the Pt nanoparticles/nanoclusters are clearly evidenced by the backscattered electrons, as they appear as white dots. An estimation of the Pt nanoparticle size was investigated by means of the ‘‘imageJ’’ software (image processing program). The size distribution was estimated between 3.8 and 13.8nm (clusters excluded) for an average value close to 7 nm. The values could be slightly over-estimated since the metallic nanoparticles appear a little bit fuzzy on the pictures. According to the literature, the optimal size would be around 3 nm.[45]

Although there are still controversies about the reduction mechanism of platinum in an atmospheric plasma discharge, and its deposition on a carbon-based surface, we suggest the tentative explanation hereunder:

(i) Energetic and reactive species from the flowing post-discharge such as Ar and O are responsible for the grafting of oxygenated functions (see Tables 1 and 2) as well as for the etching of the carbon grains. The resulting structural and chemical defects act as anchoring sites for the nanoparticles growth.

(ii) The metallic precursor is decomposed through a synergetic reaction involving the energetic species from the post-discharge, such as atomic oxygen, argon or UVs, and the temperature. Its decomposition contributes to the release of $CO_x$ volatile species. As a result, a chemical reduction of the $Pt(acac)_2$ is achieved since its Pt (+2) state decreases considerably to the benefit of the Pt (0) metallic state.

(iii) The nucleation of platinum nanoparticles is enhanced by the diffusion/convection of platinum nanoparticles on the anchoring sites. Before their fixation on the oxygen functionalities, platinum nanoparticles could be under metallic state (leading to the creation of C—O—Pt bonds) or organometallic form (since part of the precursor can still be degraded after anchoring). Indeed the previous steps of this ''one-pot process'' could occur simultaneously or successively. The Pt (+2) detected by XPS (16–23%) after plasma treatment in the open configuration could originate from (a) remaining metallic precursor, (b) interactions between platinum and the carbon black surface (Pt—O—C), and/or (c) oxidation of the platinum surface after decomposition of the organometallic precursor. Discrimination between these three chemical states of similar oxidation state is not meaningful by XPS.

(iv) Metallic nanoparticles growth on nucleation sites. Pt (0) could be present in the bulk phase and Pt (+2) on the outer surface of the particles, including bonds between carbon black and Pt in the latter case (i.e., C—O—Pt bonds).

## 3.4. Application to the PEMFC: Deposition on a Gas Diffusion Layer

Located between the membrane electrode assembly (MEA) layer and the bipolar plates (BP), the gas diffusion layers (GDL) display several functions: (i) heat transfer during the PEMFC operation; (ii) gas diffusion paths for the fuel and oxidant gases from flow channels to the catalyst surface on electrodes; and (iii) removal of the water by product out of the catalyst layer.[33] Figure 6 presents survey scans and C 1s (and Pt 4f) XPS spectra of the native GDL and of the commercial GDL, loaded with catalytic Pt nanoparticles and named here ''reference electrode.''

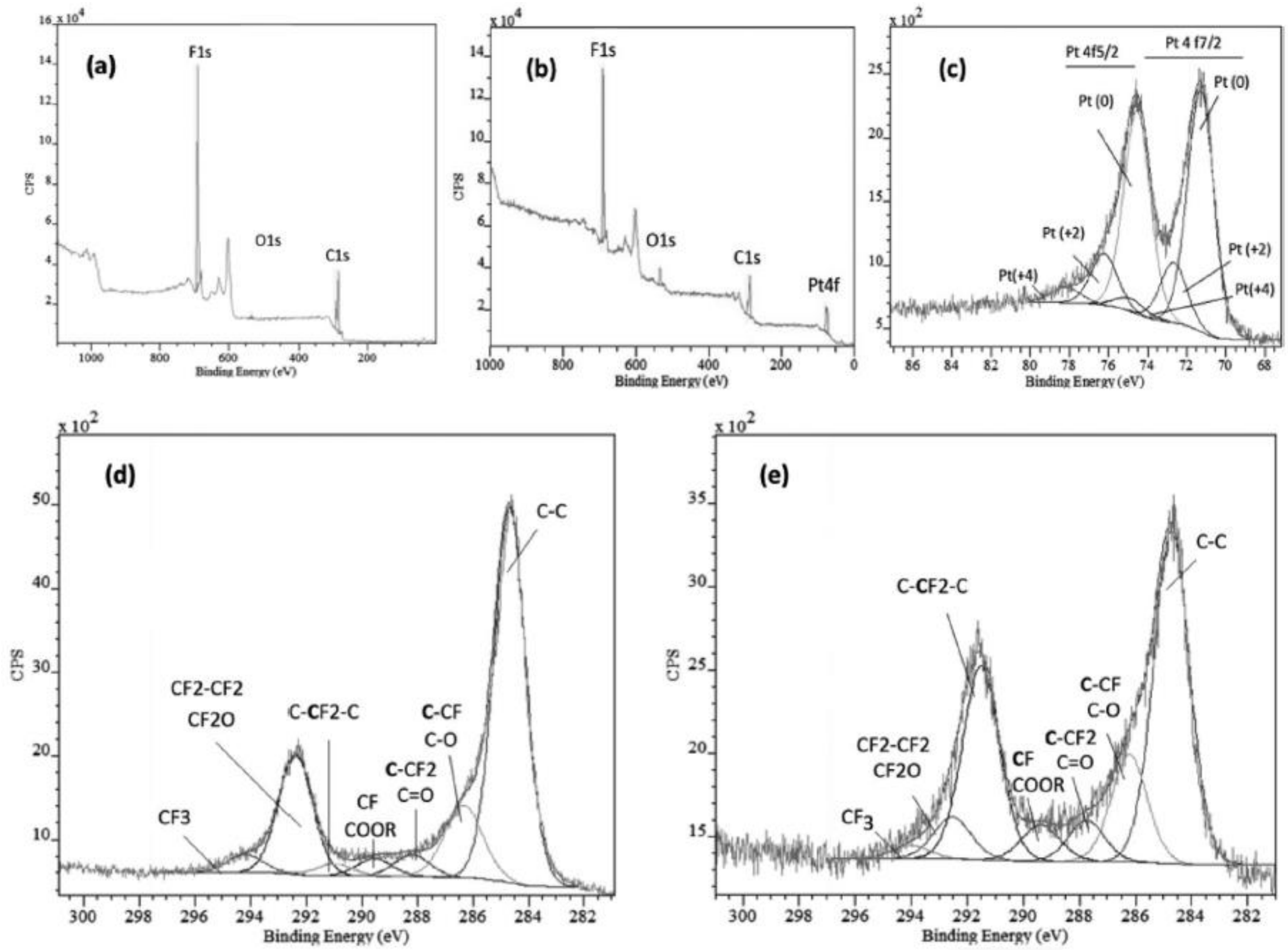

***Figure 6. Survey scans of the (a) native GDL and (b) ''reference electrode.'' Pt 4f peak of the ''reference electrode'' (c) and C 1s spectra of the native GDL (d) and ''reference electrode'' (e).***

On the native GDL, carbon, fluorine (from the presence of PTFE-based ink), and oxygen only are detected, whereas for the reference electrode, 3.9% Pt is detected by XPS, for an absolute load of 0.5 mg.cm$^{-2}$ mentioned by the manufacturer. The carbon peak exhibits a few oxygenated species, but present mostly C—C and $CF_2$ components. The fitting of the Pt peak shows that Pt is mostly under metallic form (75.5%).

The survey scans and the C1s and Pt 4f of the GDL overlaid with a CB/[Pt(acac)$_2$] mixture are given in Figure 7. The samples were treated 300 s in the open configuration by the Ar post-discharge without heating. The chemical surface composition of native GDL, reference GDL, and treated mixture overlaid on the GDL is reported in Table 5. Comparison between the results for the three systems suggest that the plasma treatment allows the formation of electrodes completely recovered by the powder mixture (as evidenced by the decrease of intensity of the fluorine peak) and containing a significant superficial amount of metallic platinum.

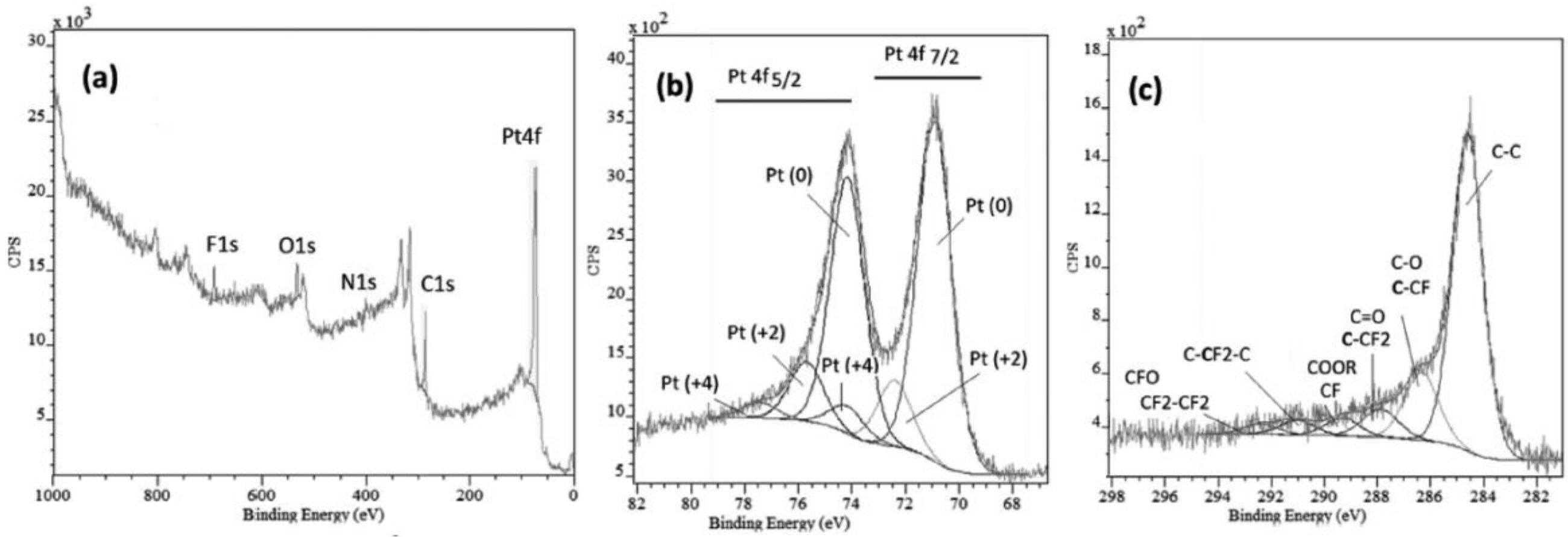

***Figure 7. Survey scan of (a) the plasma electrode (open configuration, 300 s, 60 W, 30 L min$^{-1}$ Ar), (b) Pt 4f XPS spectra, (c) C 1s XPS spectra.***

| | %Pt | %C | %N | %O | %F | %Pt (0) | Bulk Pt content [mg cm$^{-2}$] |
|---|---|---|---|---|---|---|---|
| Native paxitech GDL | – | 67.7 ± 0.8 | – | 2.1 ± 1.4 | 30.1 ± 2.2 | – | |
| Reference electrode | 3.9 ± 1.2 | 60.4 ± 5.5 | – | 8.3 ± 2.5 | 27.4 ± 7.4 | 75.5 | 0.5 |
| GDL decorated by NP by plasma | 22.1 ± 5.8 | 63.0 ± 6.1 | 1.9 ± 1.9 | 10.0 ± 1.3 | 3.0 ± 2.9 | 78.0 | 0.044 |

***Table 5. Elemental composition for native Paxitech GDL, reference electrode, and plasma electrode (60 W, 30 LPM Ar, 300 s).***

Figure 8 shows SEM pictures of the native Paxitech GDL (without catalyst), the reference electrode (GDL with catalyst), and the plasma electrode. A comparison between Figure 8c and d seems to indicate that the superficial Pt particle density is higher with the plasma method than for the reference electrode. This could be due to the specificity of the plasma approach which is more surface-sensitive than conventional impregnation techniques. This is to be compared with the bulk platinum content of 0.5 mg.cm$^{-2}$ for the reference electrode and of 0.04–0.045 mg.cm$^{-2}$ determined by ICP-MS for the plasma-loaded GDL.

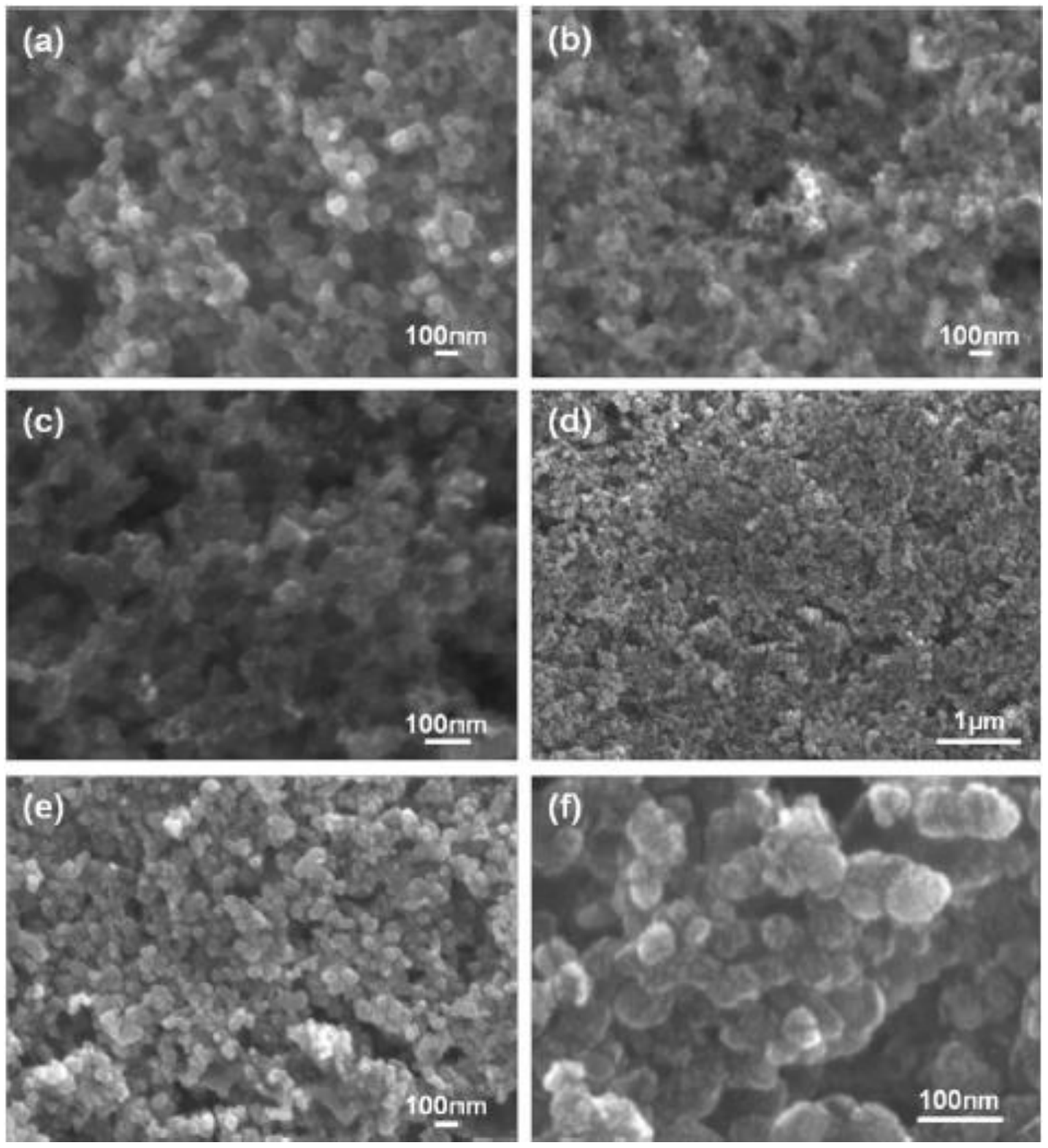

***Figure 8. SEM micrographs (secondary electrons) of the (a) native Paxitech GDL (without catalyst), (b, c) the reference electrode, and (d–f) the plasma electrode.***

The cyclic voltammetry (CV) curves of the reference Pt containing Paxitech GDL and the $Pt(acac)_2$ plasma-loaded Paxitech GDL are shown in Figure 9. While the reference electrode gives a total charge of 1000 mC, the plasma decorated GDL gives a charge of only 70 mC. However, taking into account the respective absolute Pt quantities, 0.5 and 0.044 $mg.cm^{-2}$, this leads to ECSA of 51 and 33 $m^2.g_{Pt}^{-1}$ for the reference electrode and the plasma-modified electrode, respectively. That could be explained by the higher size of the nanoparticles synthesized by plasma which present a larger size distribution. According to the method detailed in a paper of Job,[45] an ECSA of $33m^2.g_{Pt}^{-1}$ corresponds to a mean size of 8.6nm assuming that the particles are spherical and that all the Pt nanoparticles are in triple contact. According to this calculation method, nanoparticles of 3 nm could lead to an ECSA of $95m^2.g_{Pt}^{-1}$ for the reference electrode. As the value for the reference electrode is lower (51 $m^2.g_{Pt}^{-1}$), this could suggest that the catalyst particles are more accessible in the plasma-made electrode that in the reference electrode.

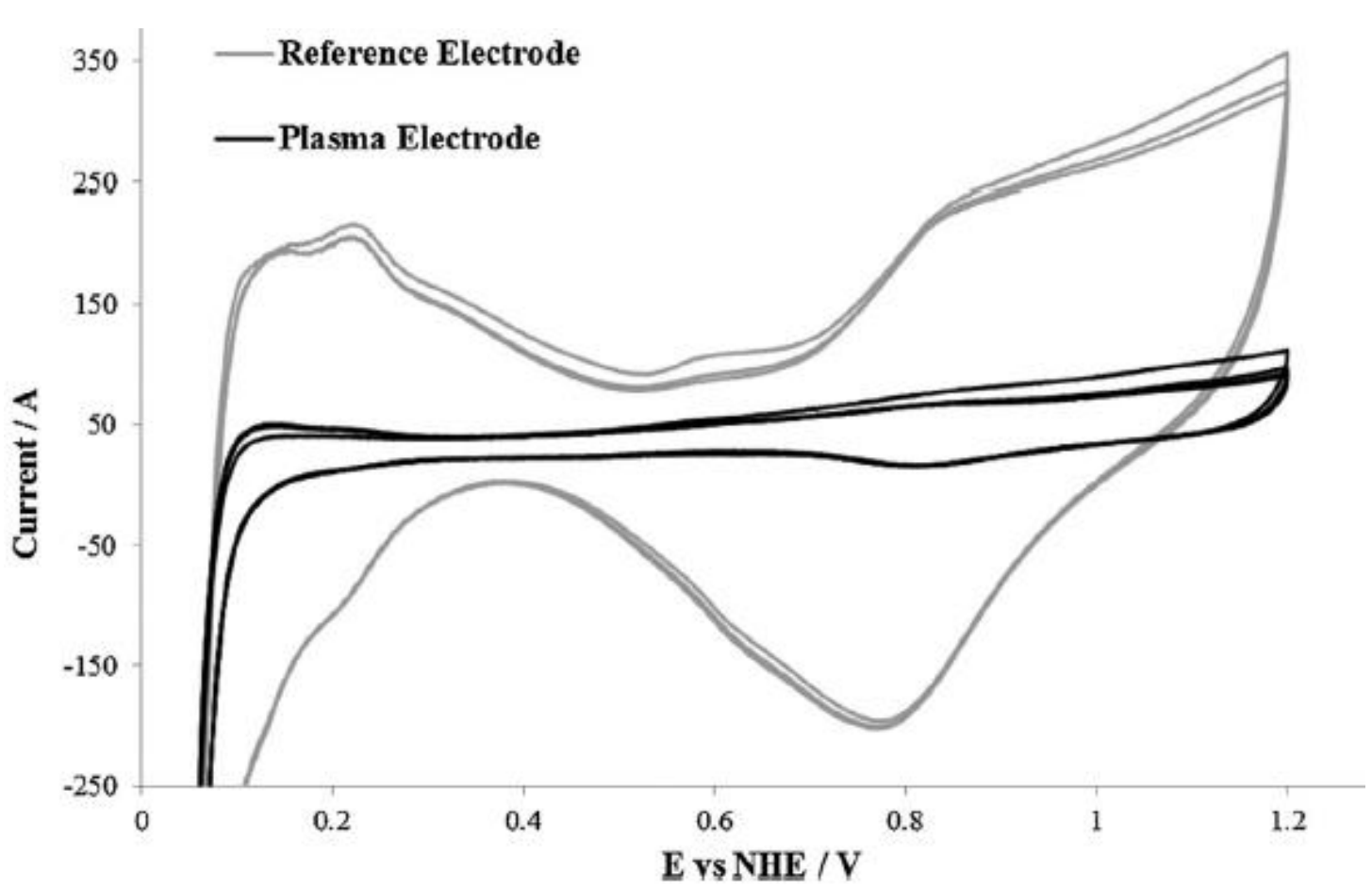

***Figure 9. Cyclic voltammogram of the single fuel cell containing the reference electrode (grey) and the plasma electrode (black) as cathode realized in the following conditions (A=25 cm², open configuration, gap=3mm, Ar=30 L.min⁻¹, 60W, 300 s).***

Figure 10 shows the evolution of the voltage for the reference electrode and the plasma-loaded electrode for various air flows. For the same current density, the voltage of the plasma-treated electrode is systematically lower than the one of the reference electrode, due to the lower platinum content of the plasma treated electrode. The losses due to diffusion, that can be detected in the area 3, are smaller for the plasma-treated electrodes than for the reference one, since less current is produced (less Pt loading). The similar slopes of the main part of the curves (area 2), reveal that losses due to the resistivity of the layer are similar. However, the losses due to electrochemical activation (area 1) are more important for the plasma-treated electrode. Consequently, the plasma electrodes exhibited an excellent contact with the membrane and so an intimate contact at the triple point. However, the size distribution and the loading have to be improved to increase the ECSA and minimize the kinetic losses (area 1).

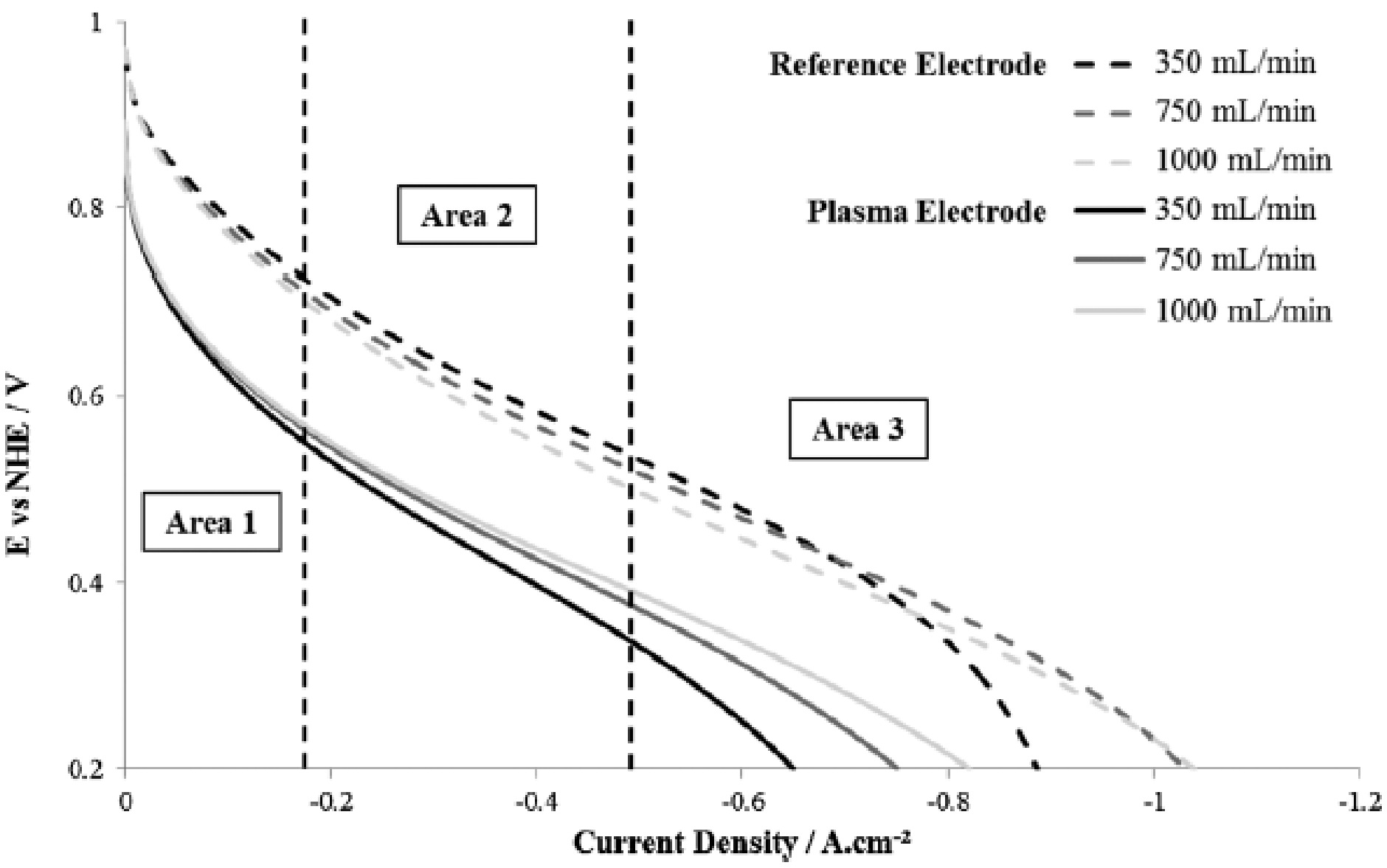

***Figure 10. Dynamical polarization curves of the mono-fuel cell containing the reference electrode and the plasma electrode as cathode realized in the following conditions (A=25cm$^2$, open configuration, gap=3 mm, Ar=30 L.min$^{-1}$, 60W, 300 s).***

# 4. Conclusion

We have showed how to synthesize and deposit platinum nanoparticles on carbon supports from a mixture of carbon black with [Pt(acac)$_2$] by using an atmospheric plasma treatment. Contrarily to the usual processes from the literature, the technique is simple, fast, robust, and operates in one single step at a temperature close to the ambient temperature. According to our results, it leads to the following conclusions: (i) although heating the sample was not a necessary condition, it improves the amount of Pt grafted. Indeed, for a same plasma treatment, the Pt grafted is 18% at 383 K while 10% at 333 K; (ii) the plasma treatment could be necessary to deposit metallic particles by decomposing the organics ligands below the decomposition temperature of the [Pt(acac)$_2$] on the surface of the carbon powder; (iii) for a treatment time increasing from 30 to 600 s, the Pt content detected on the surface increased from 6 (mostly under its metallic form) to 13%; (iv) the best two configurations to obtain Pt contents as high as 11% are either the confined configuration at 10 mm or the open configuration at 3mm. The advantage of the latter is the ability to treat larger carbon substrates areas.

For ‘‘more targeted’’ applications, the plasma treatment of the carbon black mixture overlaid on a gas diffusion layer enables to obtain electrodes for PEMFC tested in a single cell. Although the ECSA ($33 m^2.g_{Pt}^{-1}$) of the plasma-made electrode is lower than the one of the reference electrode ($51 m^2.g_{Pt}^{-1}$), it must be taken into account that the absolute Pt content of the plasma-electrodes determined by ICP is much smaller than the one of the reference electrode (0.044 compared to 0.5 $mg.cm^{-2}$). This suggests a good contact between the plasma electrode and the membrane hence an intimate contact between the catalyst/membrane/arrival of the reactants where the redox reactions take place in the cell. The size distribution and the Pt loading have to be improved in an outlook.

# 5. Acknowledgements

This work was financially supported by the Walloon Region (INNOPEM project, grant no. 1117490 and Hylife project, grant no 141 01 35) and partially by the Science Policy Inter-University Attractive Pole (IUAP) ‘‘Physical Chemistry of plasma surface interactions’’ (IAP-VII/12, P7/34). N. Job thanks the Fonds de Bay for funding. V.D. thanks the IUAP ‘‘Planet Topers’’ for instrumentation and the F.R.S.-FNRS and the ERC Starting Grant ‘‘ISoSyc’’ for support.